\documentclass[twocolumn]{aastex62}
\usepackage{amsmath}
\usepackage{float}
\usepackage{afterpage}
\graphicspath{{./}{figures/}}

\shorttitle{Observational Signatures of Sub-Relativistic Meteors}
\shortauthors{Siraj \& Loeb}

\begin{document}

\title{Observational Signatures of Sub-Relativistic Meteors}

\email{amir.siraj@cfa.harvard.edu, aloeb@cfa.harvard.edu}

\author{Amir Siraj}
\affil{Department of Astronomy, Harvard University, 60 Garden Street, Cambridge, MA 02138, USA}

\author{Abraham Loeb}
\affiliation{Department of Astronomy, Harvard University, 60 Garden Street, Cambridge, MA 02138, USA}



\begin{abstract}

It is currently unknown whether solid particles larger than dust from supernova ejecta rain down on Earth at high speeds. We develop a hydrodynamic and radiative model to explore the detectability of $\gtrsim 1 \mathrm{\; mm}$ sub-relativistic meteors. We find that a large fraction of the meteor energy during its passage through the Earth's upper atmosphere powers the adiabatic expansion of a hot plasma cylinder, giving rise to acoustic shocks detectable by infrasound microphones. Additionally, a global network of several hundred all-sky optical cameras with a time resolution of $\lesssim 10^{-4} \mathrm{\;s}$ would be capable of detecting $\gtrsim 1 \mathrm{\; mm}$ sub-relativistic meteors.

\end{abstract}



\section{Introduction}
The study of meteors is well-established \citep{1998SSRv...84..327C} but attention is focused on speeds $\sim 10^{-4} \mathrm{\: c}$. Objects moving much faster could potentially have qualitatively different signatures and therefore could be missed by current searches. In this paper, we aim to explore those signatures so as to design the best detection strategies for them.

Empirical evidence indicates that at least one nearby supernovae resulted in the $^{60}\mathrm{Fe}$ and other radionuclides detected in deep-ocean samples \citep{1999PhRvL..83...18K, 2004PhRvL..93q1103K, 2012PASA...29..109F, 2016Natur.532...69W}, the lunar surface \citep{2016PhRvL.116o1104F}, and cosmic rays \citep{1974Sci...184.1079R, 2015PhRvL.115r1103K, 2018PhRvD..97f3011K}.

\cite{1949PhRv...76..583S} showed that dust grains can be accelerated to sub-relativistic and relativistic speeds due to radiation pressure from supernovae.  \cite{1972Ap&SS..16..238H} proposed that dust grains may be a source of the phenomenology attributed to particle showers of ultrahigh energy cosmic rays (UHECRs). Processes affecting the origin and survival of relativistic dust have been extensively studied \citep{1956Tell....8..268H, 1973Ap&SS..21..475B, 1977ICRC....2..358B, 1977Ap.....13..432E, 1993Ap&SS.205..355M, 1999APh....12...35B}, and \cite{2015ApJ...806..255H} showed that grains in the Milky Way Galaxy can be accelerated nearly to the speed of light with Lorentz factor $\gamma < 2$. The origin and nature of UHECRs still remain unclear \citep{2019BAAS...51c..93S}.

In contrast with cosmic ray events, an object moving through the atmosphere with size and speed such that air molecules have time to share energy will cause a hydrodynamic shock wave in the surrounding gas instead of a particle shower.

Significant slow-down for objects traveling through the interstellar medium (ISM) occurs at distance $d_{ISM}$ when the accumulated ISM mass is comparable to the mass of the object \citep{2018ApJ...868L...1B},

\begin{equation}
    d_{ISM} \sim \frac{r \rho_{obj}}{m_p n_p} \; \; ,
\end{equation}
where $m_p$ is the proton mass, $n_p$ is the mean proton number density of the ISM, taken to be $\sim 1 \; \mathrm{cm^{-3}}$, $r$ is the radius of the object, and $\rho_{obj}$ is the mass density of the object. For an object of size $r \sim 1 \mathrm{\; mm}$ and mass density $\rho_{obj} \sim 3 \mathrm{\;g \; cm^{-3}}$, the slow-down distance $d_{ISM} \gtrsim \mathrm{50 \; kpc}$. Therefore, if small solid-density clumps $\gtrsim 1 \mathrm{\; mm}$ are expelled from supernovae in the Galaxy, they could appear in the Earth's atmosphere at their initial sub-relativistic speeds of $\sim 0.1 \mathrm{\;c}$ \citep{Weiler2003}.

The mass fraction of supernova ejecta with size $\gtrsim r$, $\eta_r$, that causes a flux of $\dot{n}_r$ such objects at Earth is,

\begin{equation}
    \eta_r = \left( \frac{16 \pi}{3} \right) \left( \frac{\dot{n}_r r^3 \rho_{obj} \: d_{SN}^2}{\dot{n}_{SN} M_{SN} R_{\oplus}^2} \right) \; \; ,
\end{equation}
where $d_{SN}$ is the distance to the supernova, $\dot{n}_{SN}$ is the supernova rate, $M_{SN}$ is the ejected mass in solids from the supernova, and $R_{\oplus}$ is the Earth's radius. \cite{2007ESASP.622..117S} found the rate of supernovae in the Galactic Center to be $\dot{n}_{SN} \sim 0.02 \mathrm{\; yr^{-1}}$, and \cite{2009ApJ...703..642C} estimated that each supernova releases $M_{SN} \sim 0.1 \; M_{\odot}$ of dust. Given these estimates, the aforementioned mass and mass density values, and the distance to the Galactic center $d_{SN} \sim 8 \mathrm{\; kpc}$, we estimate the fraction of dust mass from each supernova with size $\gtrsim 1 \mathrm{\; mm}$ that results in a flux of 1 such meteor per month on Earth to be a modest $\eta_{1\mathrm{mm}} \sim 10^{-4}$, corresponding just to roughly an Earth mass per supernova. Dense clumps, or "bullets," have been identified in supernova remnants, serving as evidence for clumpiness at low density that could lead to fragmentation into the small solid-density objects we consider \citep{2002ApJ...574..155W}.

Since the atmospheric scale height on Earth is $\sim 8 \mathrm{\; km}$, our hydrodynamic simulations focus on the last $10 \mathrm{\; km}$ of an object's path in the atmosphere, where $70 \%$ of the air mass is encountered. For simplicity, we consider trajectories normal to the Earth's surface. 

In what follows, we explore observational signatures of potential sub-relativistic meteors. In Section \ref{conditions}, we derive size and speed constraints such that a hydrodynamic model applies. Next, we describe our hydrodynamic model in Section \ref{simulation}. We then report the results of our hydrodyanmic simulations in Section \ref{results}. Finally, we discuss our main conclusions in Section \ref{discussion}.

\section{Conditions for a Hydrodynamic Shock}
\label{conditions}
Adopting an atmospheric density at an altitude, $z$, of $\rho_{air}(z) = 10^{-3} \; \mathrm{g \; cm^{-3}} \: e^{-z/\mathrm{8 \; km}}$ and an object density of $\rho_{obj} \sim 3 \mathrm{\;g \; cm^{-3}}$, the altitude at which the traversed air mass is comparable to the mass of the object (approximately the altitude at which a meteor appears), $z_{min}$, can be expressed as a function of object radius, $r$:

\begin{equation}
    z_{min} (r) \sim 8 \; \mathrm{km} \: \ln({200 \; \mathrm{cm}/r}) \; \; .
\end{equation}

For a hydrodynamic shock to form in the atmosphere, the slow-down timescale of the object, $\tau_{slow} = (10 \mathrm{\; km}/v)$ where $v$ is the speed of the object, must be longer than the collision time scale between molecules in the ambient air at altitude $z$, $\tau_{coll} = (n_z \sigma v_{th})^{-1}$, where $n_z \sim 10^{19} \: \mathrm{cm^{-3}} \: e^{-z/\mathrm{8\;km}}$ is the number density, $\sigma \sim 10^{-15} \mathrm{\; cm^2}$, is the collision cross section, and $v_{th} \sim 3 \times 10^4 \mathrm{cm \; s^{-1}}$ is the thermal speed, yielding,

\begin{equation}
      \beta \; e^{z/\mathrm{8\;km}} \lesssim 10^{14}  \; \; ,
      \label{eq:time}
\end{equation}
where $\beta = v/c$. The surrounding gas acts as a fluid where particles share their energy with each other and behave collectively through their pressure and temperature. Condition (\ref{eq:time}) does not place any meaningful constraint on the object's speed unless the assumption that the object slows down over $10 \; \mathrm{km}$ is invalid. 

We therefore consider the penetration depth of incoming air particles into the object. In the rest frame of a dust particle, the incoming air particles strike it at a speed $v$. If they can penetrate through the entire object, they will break it up and not just evaporate the outer layer, so the object would lose its integrity very quickly and get dispersed, broadening the cross-sectional area of its interaction with air and slowing down much more quickly anticipated by our estimate of $\tau_{coll}$ used in condition (\ref{eq:time}).

We use the penetration depth of air (mostly nitrogen) nuclei into a solid to set a lower limit on the object size as a function of speed. We compute this depth as a function of speed, by dividing the kinetic energy of a nitrogen ion at a given speed, $v$, namely $E_N =(1/2) m_N v^2$ (where $m_N$ is the nitrogen mass), and divide by the energy loss per unit length, $dE/dx$ (\citealp{2017ApJ...837....5H}, Figure 2), finding the penetration depth, $d_{pen}$, as a function of speed to be,

\begin{equation}
    d_{pen}(\beta) \approx 85 \; \mathrm{cm} \:\beta^{3.2}, \; \;0.025 < \beta < 0.45 \; \; .
\end{equation}
The penetration depth condition is then, 
\begin{equation}
    d_{pen}(\beta) < r \; \; .
    \label{eq:pendepth}
\end{equation}
We assume that when condition (\ref{eq:pendepth}) is violated, the object is slowed down and dispersed instantaneously and our hydrodynamic model does not apply.

Finally, our hydrodynamics shock wave model applies to distances longer than the mean free path of the surrounding air particles, $l = (n_z \sigma)^{-1}$, because it is the minimum scale on which the air particles share their energy and behave collectively as a fluid,

\begin{equation}
    r > l \; \; .
    \label{eq:mfp}
\end{equation}
Since the mean free path, $l$, is a function of altitude, $z_{min}$, which is, in turn, a function of object radius $r$, we find that $r > l$ corresponds to $r \gtrsim 0.9 \mathrm{\; mm}$.

\begin{figure}
  \centering
  \includegraphics[width=0.9\linewidth]{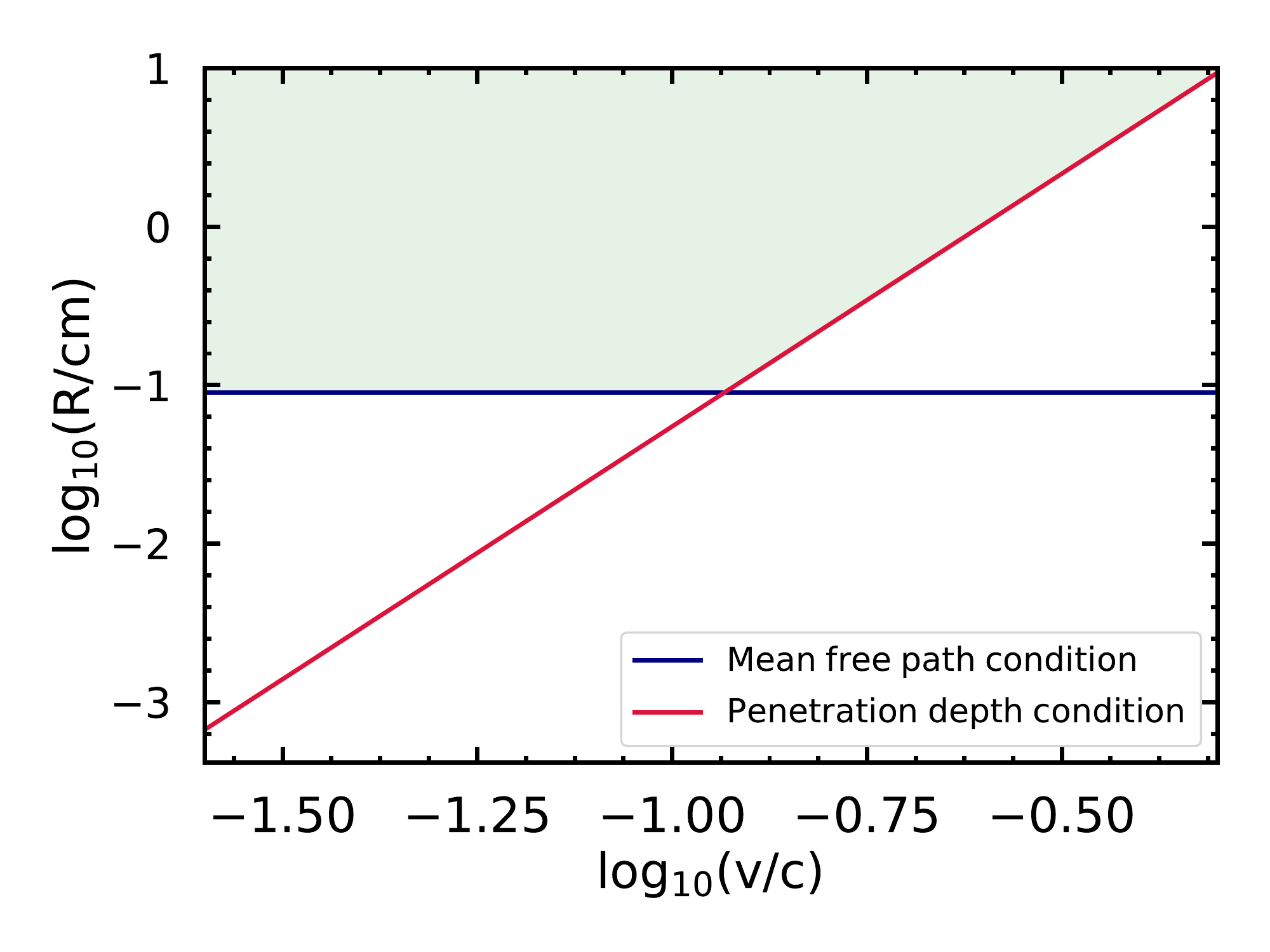}
    \caption{Parameter space of object radius, $r$, and speed, $v$, for which conditions (\ref{eq:pendepth}) and (\ref{eq:mfp}) are met and our hydrodynamic approach applies is shaded.}
    \label{fig:hydro}
\end{figure}

Conditions (\ref{eq:pendepth}) and (\ref{eq:mfp}) are therefore dominant for ensuring that the air behaves as a fluid, and the parameter space for which our hydrodynamic approach applies is shaded in Figure \ref{fig:hydro}. 

\section{Hydrodynamic Simulation}
\label{simulation}

A sub-relativistic meteor in the aforementioned parameter space does not lead to a synchronous cylindrical explosion, since the adiabatic cooling time is significantly shorter than the slow-down time. However, because the object typically moves much faster than the shock by a few orders of magnitude, locally the cylindrical approximation is appropriate. 

Adopting a time resolution\footnote{The gas cools from its initial temperature to several $\times \: 10^6 \mathrm{\; K}$ on a timescale of order $10^{-9} \mathrm{\; s}$.} of $10^{-9} \mathrm{\; s}$ for a timescale such that each cylindrical segment of gas cools to the first ionization energy of nitrogen, $\sim 10^5 \mathrm{\; K}$, for a meteor traveling at $0.1 \; c$,\footnote{$\sim 1.5 \times 10^{-7} \mathrm{\; s}$, $\sim 8 \times 10^{-7} \mathrm{\; s}$, and $\sim 1.5 \times 10^{-6} \mathrm{\; s}$, for meteors of size $r = 1 \mathrm{\;mm}$, $1 \mathrm{\;cm}$, and $10 \mathrm{\;cm}$, respectively.} we divide the final $10 \; \mathrm{km}$ of the meteor's trajectory into $\sim 3.3 \times 10^4 - 3.3 \times 10^5$ cylindrical segments, each with length $30 - 300 \; \mathrm{cm}$ and radius $r$. Since the conditions inside any two segments would only vary significantly if separated by a distance of order an atmospheric scale height, we simulate 100 of the cylindrical segments, evenly spaced throughout the $10 \; \mathrm{km}$, and apply the results to the $\sim 3.3 \times 10^3 - 3.3 \times 10^4$ local segments.

For each segment, the initial energy deposited is proportional to the local air density and such that the total energy deposited over the entire $10 \mathrm{\; km}$ is $70 \%$ of the object's initial kinetic energy. This gives rise to an initial temperature, $T$, from the energy being shared among all air particles (now ionized) in the segment as well as the fraction of the object's initial mass deposited in the segment, with a collective mass density, $\rho$. This generates an initial pressure in the gas that drives a shock with an enhanced density $\bar{\rho}$ \citep{Richardson2019}, 

\begin{equation}
    \frac{\bar{\rho}}{\rho} = \frac{4 M^2}{M^2 + 3} \; \; ,
\end{equation}
having adopted an adiabatic index of $\gamma = 5/3$, where $M$ is the Mach number associated with the ratio of interior to exterior temperatures, $\zeta = T/T_{ext}$,

\begin{equation}
   M = \sqrt{\frac{4\sqrt{4\zeta^2 - 7\zeta + 4} + 8\zeta - 7}{5}} \; \; .
\end{equation}
The mass inside the segment is concentrated into a shell with density, $\bar{\rho}$, and fractional thickness, $\rho/\bar{\rho}$. This leads to an updated initial temperature, $T$.

The simulation then works as follows. For each timestep, the temperature, $T$, leads to an updated shell density, $\bar{\rho}$, as well as an updated shell growth rate, $\dot{R}$. Simultaneously, the shell loses energy to bremmstrahlung (since the emitting ionized gas is optically-thin) and adiabatic expansion. At the end of each timestep, the shell's energy, $E$, is used to recalculate the temperature, $T$, for the start of the following timestep. As the segment grows, it accumulates more air mass that is added to the mass of the shell.

During each timestep, the shell with density $\bar{\rho}$ expands radially at the rate, $\dot{R}$,

\begin{equation}
    \dot{R} = v_c M \; \;,
\end{equation}
where $v_c$ is the sound speed in the shell,

\begin{equation}
    v_c = \sqrt{\frac{5 P}{3 \bar{\rho}}} \; \; ,
\end{equation}
and where $P = n_N k T$ is the gas pressure, with nitrogen ion density (roughly uniform inside the shell), $n_N = 4.3 \times 10^{19}\: \mathrm{cm^{-3}} \; (\bar{\rho}/\rho) \: e^{-z/\mathrm{8 \; km}}$. The gas cools by bremmstrahlung at the frequency-integrated rate $\dot{E}_{B}$ 
\citep{1979rpa..book.....R},

\begin{equation}
    \dot{E}_{B} = 1.4 \times 10^{-27} \left( \frac{T}{\mathrm{K}} \right)^{1/2} \left( \frac{n_e n_N}{\mathrm{cm^{-6}}} \right) \left( \frac{V(T)}{\mathrm{cm^3}} \right) Z^2 \bar{g}_B \; \mathrm{erg \; s^{-1}} \; \; ,
\end{equation}
where $n_e$ is the number density of electrons, and of ions, $Z$ is the ion charge in units of electron charge, $\bar{g}_B \sim 1.2$ is the velocity-averaged Gaunt factor, and $V(T) = \pi (R^2 - R^2\rho^2/\bar{\rho}^2)l$ is the volume of the shell, where $l$ is the  length of the segment. The first ionization energy of nitrogen is $14.5 \; \mathrm{eV}$, so $E \sim \frac{3}{2} n k T$ yields a temperature of $T \sim 10^5 \; \mathrm{K}$ at which air is fully singly ionized. We follow the same process for the subsequent ionization energies of nitrogen, and at each temperature treat the air as fully ionized above the relevant energy cutoff for a certain ionization, deriving $n_e$ as a integer factor of $0 - 7$ of $n_N$ at each timestep.

\begin{figure*}[!th]
  \centering
  \includegraphics[width=0.8\linewidth]{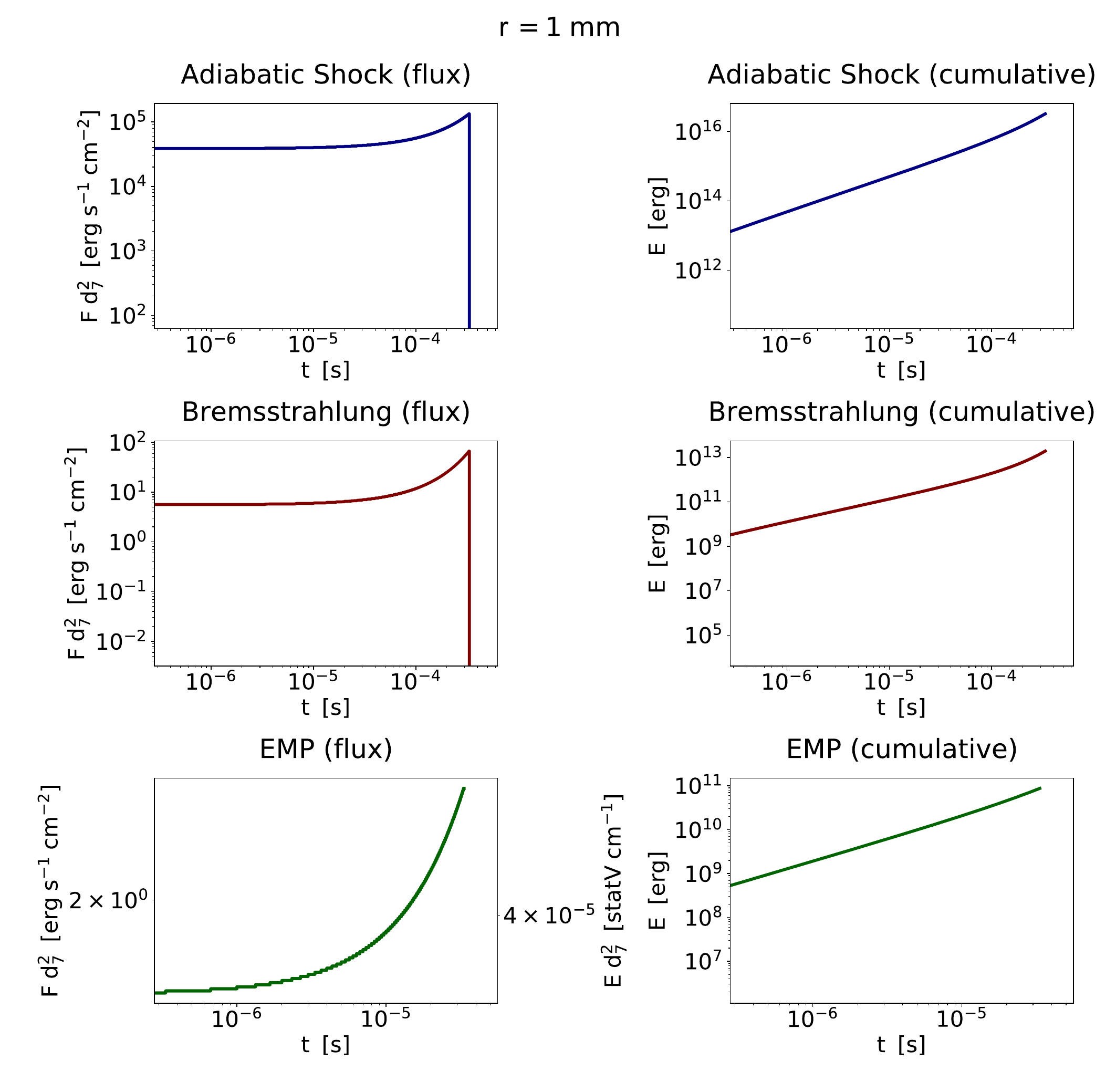}
    \caption{Adiabatic shock, bremsstrahlung, and electromagnetic pulse for a $r = 1\; \mathrm{mm}$, $\beta = 0.1$ meteor. From upper left: the adiabatic shock flux for a $1 \mathrm{\; cm}$ detector located $100 \mathrm{\; km}$ away from the explosion, the total cumulative adiabatic shock, the bremsstrahlung radiation flux for a $1 \mathrm{\; cm}$ detector located $100 \mathrm{\; km}$ away from the explosion, the total cumulative bremsstrahlung radiation, the EMP energy flux for a $1 \mathrm{\; cm}$ detector located $100 \mathrm{\; km}$ away from the explosion (as well as the corresponding electric field), and the total cumulative EMP energy.}
    \label{fig:emission1mm}
\end{figure*}

\begin{figure*}[!th]
  \centering
  \includegraphics[width=0.85\linewidth]{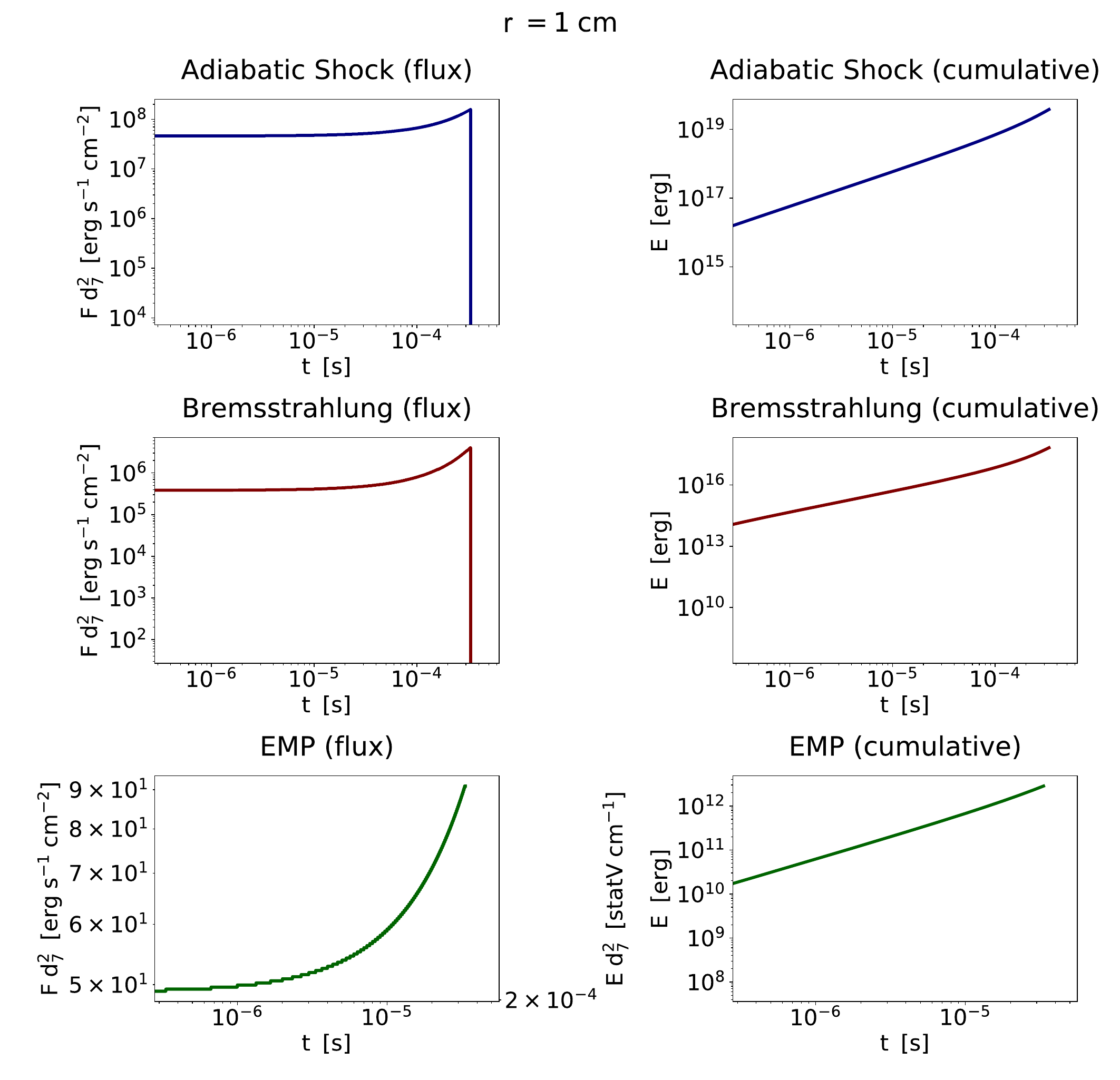}
    \caption{Same as Figure \ref{fig:emission1mm} for $r = 1\; \mathrm{cm}$.}
    \label{fig:emission1cm}
\end{figure*}

\begin{figure*}[!th]
  \centering
  \includegraphics[width=0.85\linewidth]{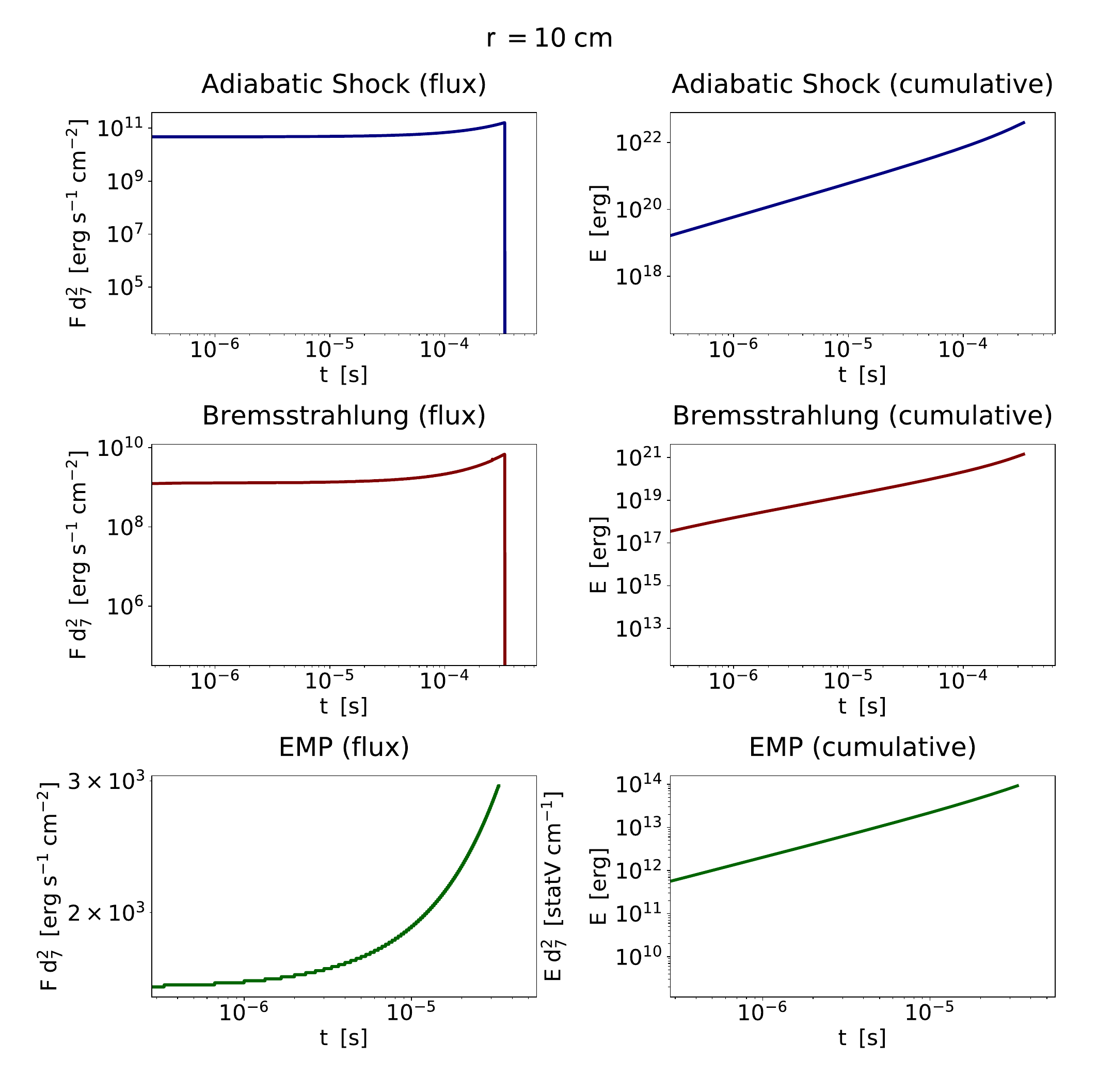}
    \caption{Same as Figure \ref{fig:emission1mm} for $r = 10\; \mathrm{cm}$.}
    \label{fig:emission10cm}
\end{figure*}

\begin{figure*}[!th]
  \centering
  \includegraphics[width=0.85\linewidth]{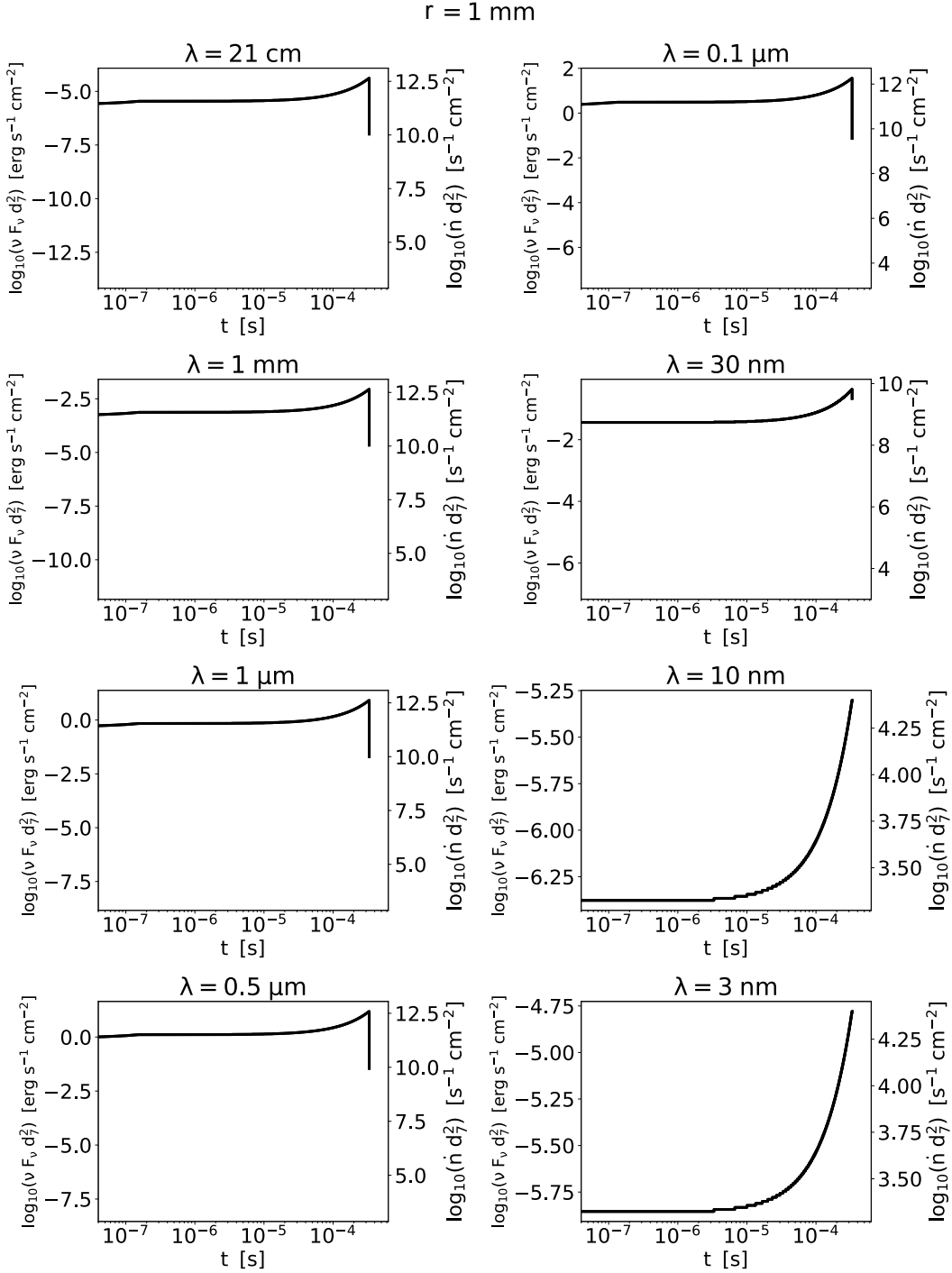}
    \caption{Bremsstrahlung radiation flux reaching a $1 \mathrm{\; cm}$ detector located $100 \mathrm{\; km}$ away from the explosion, at wavelengths $\lambda = 21 \mathrm{\;cm}$, $1 \mathrm{\;mm}$, $1 \mathrm{\;\mu m}$, $0.5 \mathrm{\;\mu m}$, $0.1 \mathrm{\;\mu m}$, $30 \mathrm{\;nm}$, $10 \mathrm{\;nm}$, and $3 \mathrm{\;nm}$, for a $r = 1\; \mathrm{mm}$, $\beta = 0.1$ meteor.}
    \label{fig:ff1mm}
\end{figure*}

\begin{figure*}[!th]
  \centering
  \includegraphics[width=0.8\linewidth]{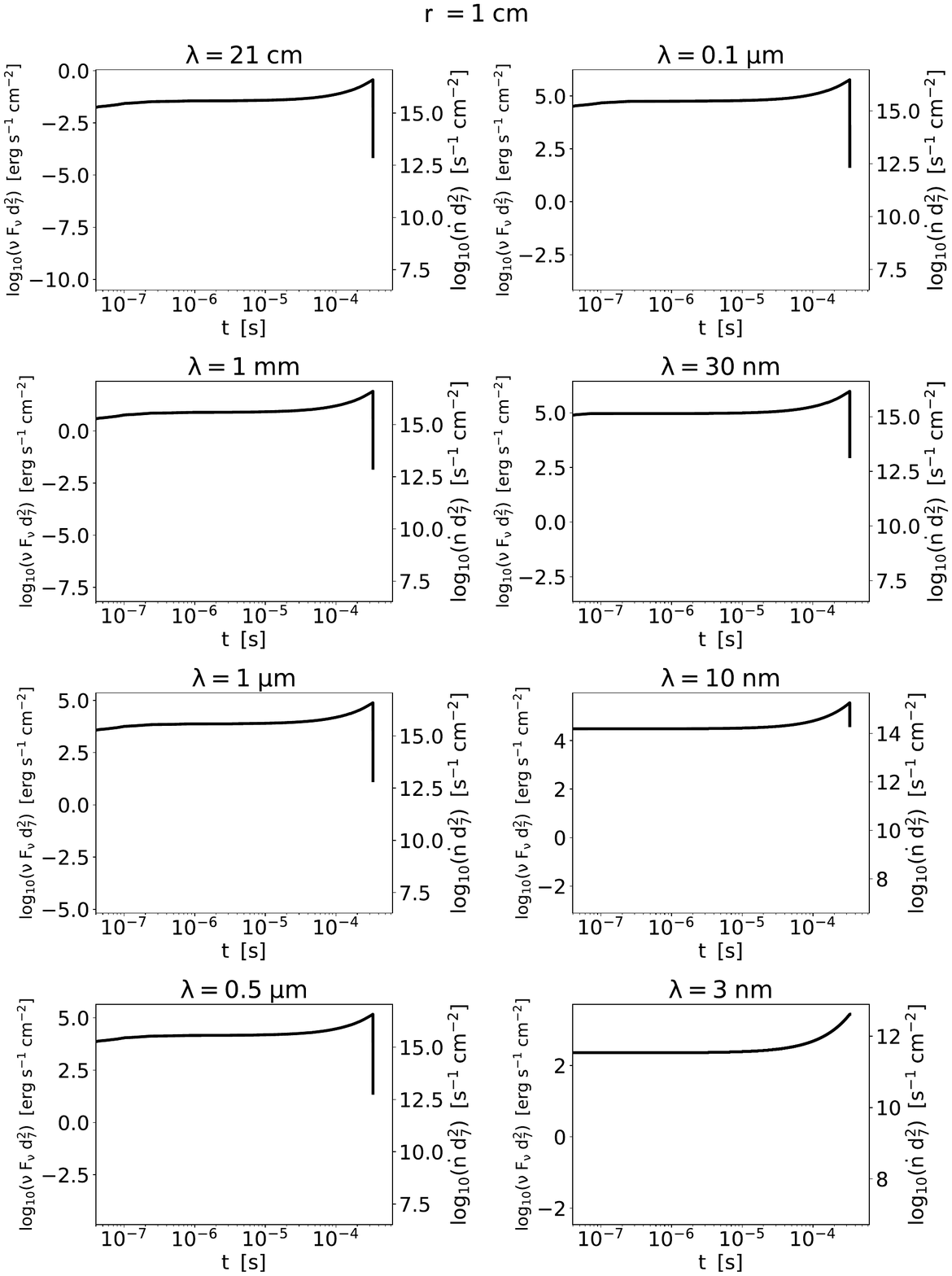}
    \caption{Same as Figure \ref{fig:ff1mm} for $r = 1\; \mathrm{cm}$.}
    \label{fig:ff1cm}
\end{figure*}

\begin{figure*}[!th]
  \centering
  \includegraphics[width=0.8\linewidth]{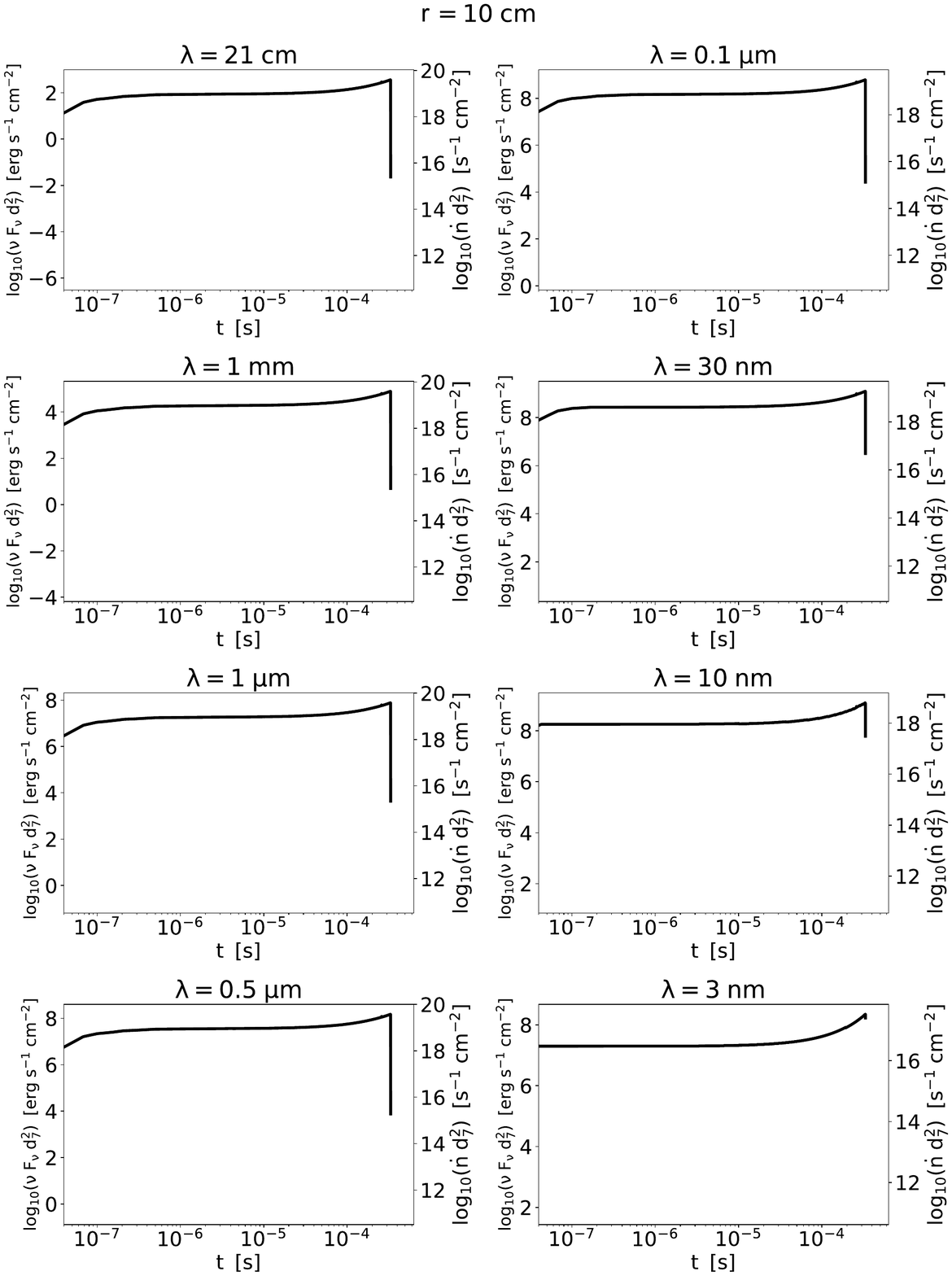}
    \caption{Same as Figure \ref{fig:ff1mm} for $r = 10\; \mathrm{cm}$.}
    \label{fig:ff10cm}
\end{figure*}

The gas also cools by adiabatic cooling at the rate, $\dot{E}_{ad}$,

\begin{equation}
    \dot{E}_{ad} = P \: V(T) \:\dot{V}(T) \; \; ,
\end{equation}
leading to an updated energy, $E$, which is used to calculate the updated temperature, $T$, at the end of each timestep through the relation,

\begin{equation}
    E = U(T) \: V(T) \; \; ,
\end{equation}
where $U(T) = \frac{3}{2} nkT$ is the energy density of the shell.

We also calculate bremsstrahlung rates for wavelengths of $\lambda = 21 \mathrm{\;cm}$, $1 \mathrm{\;mm}$, $1 \mathrm{\;\mu m}$, $0.5 \mathrm{\;\mu m}$, $0.1 \mathrm{\;\mu m}$, $30 \mathrm{\;nm}$, $10 \mathrm{\;nm}$, and $3 \mathrm{\;nm}$, with $v = (c/\lambda)$, based on the emissivity, 

\begin{equation}
\begin{aligned}
    & \dot{E}_{B}^{\nu} = 6.8 \times 10^{-38} \left(\frac{T}{\mathrm{K}} \right)^{1/2} \left( \frac{n_e n_N}{\mathrm{cm^{-6}}} \right) \\ 
    & Z^2 e^{-h\nu/kT} \bar{g}_B \left(\frac{V(T)}{\mathrm{cm^{3}}}\right) \; \mathrm{erg \; s^{-1} \; Hz^{-1}} \; \; ,
\end{aligned}
\end{equation}
which applies to all frequencies below the cutoff frequency, $\nu_{cutoff} \approx kT/h$. We apply the above treatment to all segments and sum the resultant quantities with the appropriate time delay.

In addition, we calculate the resulting electromagnetic pulse (EMP) from the charge separations of the electrons. For simplicity, we ignore magnetic fields. Electrons slow down through collisions at their mean-free-path, $\lambda_M$, but also through electrostatic effects over a distance of order the Debye length, $\lambda_D$. In the air plasmas under consideration here, $\lambda_D < \lambda_M$, and so we use the Debye length as the maximum charge separation distance of electrons for maintaining global quasi-neutrality. The thermal electron speed is, 

\begin{equation}
    v_T^e \approx \sqrt{3kT/m_e} \; \;,
\end{equation}
and the Debye length is,

\begin{equation}
   \lambda_D = 7.43 \times 10^2 \left(\frac{T}{\mathrm{K}}\right)^{1/2} \left(\frac{n_e}{\mathrm{cm^{-3}}}\right)^{-1/2} \; \mathrm{cm} \; \; .
\end{equation}
The built-up electric field that is released over the timescale $\lambda_D/v_T^e$ is then,

\begin{equation}
    E_{EMP} = \frac{m_e v_e^2}{2 d_D e} \; \;,
\end{equation}
where $e$ is the electron charge. We calculate the EMP for each segment and sum over segments with the appropriate time delay.

\section{Results}
\label{results}

We apply our approach to meteors with radii $r = 1 \mathrm{\;mm}$, $1 \mathrm{\;cm}$, and $10 \mathrm{\;mm}$, traveling at $\beta = 0.1$, as a fiducial example.

The resulting energy fluxes as a function of time (as well as a cumulative energy) for the adiabatic shock, bremsstrahlung, and EMP are indicated in Figures \ref{fig:emission1mm} - \ref{fig:emission10cm}. The bremsstrahlung fluxes as a function of time for the eight previously indicated wavelengths are indicated in Figures (\ref{fig:ff1mm} - \ref{fig:ff10cm}). The fraction of energy released in the adiabatic shock are $\sim 1$, $\sim 0.98$, and $\sim 0.96$, in order of increasing size. The fraction of energy released in the bremsstrahlung radiation are $\sim 5 \times 10^{-4}$, $\sim 0.02$, and $\sim 0.04$, in order of increasing size. The fraction of energy released in the EMP are $\sim 2 \times 10^{-5}$, $\sim 7 \times 10^{-6}$, and $\sim 2 \times 10^{-9}$, in order of increasing size.

The bremsstrahlung signals peak at UV wavelength of $\sim 0.1 \; \mathrm{\mu m}$ for $r = 1 \mathrm{\;mm}$, and at $\sim 30 \mathrm{\;nm}$ for $1 \mathrm{\;cm}$ and $10 \mathrm{\;cm}$ meteors, decreasing at longer wavelengths. Since UV radiation is absorbed in the atmosphere, the best observational window is in the optical-infrared bands, where the atmosphere transmits the emitted light. The peak efficiencies are all of order several $\times 10^{-1}$ of the total bremsstrahlung radiation. The $\sim 0.5 \; \mathrm{\mu m}$ (optical) efficiences, in increasing size order, are $\sim 0.2$, $\sim 0.04$, and $\sim 0.02$ of the total bremsstrahlung radiation, corresponding to $\sim 10^{9}$, $\sim 10^{13}$, and $\sim 10^{16}$ photons reaching a $\sim 1 \mathrm{\; cm^2}$ optical ground detector at a distance of $\sim 100 \mathrm{\; km}$ over $\sim 10^{-4} \; \mathrm{s}$ for meteors of size $r = 1 \mathrm{\;mm}$, $1 \mathrm{\;cm}$, and $10 \mathrm{\;mm}$, respectively.

\section{Discussion}
\label{discussion}

The results suggest that infrasound microphones \citep{2008JGRD..11312115L} with directional sensors searching for acoustic shocks lasting for $\sim 10^{-4} \mathrm{\; s}$ originating from a range of altitudes of order the atmospheric scale height could be an effective way to search for sub-relativistic meteors.

Searches for optical flashes lasting for $\sim 10^{-4} \mathrm{\; s}$ originating from a range of altitudes of order the atmospheric scale height should serve as an effective detection method. Our results indicate that $\sim 1 \mathrm{\; cm^2}$ optical detector with a time resolution of $\lesssim 10^{-4} \; \mathrm{s}$ could easily detect a $\sim 1 \; \mathrm{mm}$ sub-relativistic meteor out to a distance of $\sim 10^3 \mathrm{\;km}$. A global network of $\sim 600$ such detectors with all-sky coverage \citep{2019arXiv190603270S} could detect a few sub-relativistic meteors per year if the fraction of supernova dust contained in objects of size $\gtrsim 1 \mathrm{\; mm}$, is $\eta_{1\mathrm{mm}} > 10^{-4}$.

Since sub-relativistic $\sim 1 \; \mathrm{cm}$ meteors should radiate $\sim 2\%$ ($10^{18} \mathrm{\; erg}$) of their kinetic energy, such a flash would be approximately an order of magnitude more energetic than the least energetic fireballs reported in the US Government's CNEOS database.\footnote{https://cneos.jpl.nasa.gov/fireballs/} If the time resolution of sensors the CNEOS network is $\lesssim 10^{-4} \; \mathrm{s}$, then CNEOS could provide an optimal dataset in which to search for sub-relativistic $r \sim 1 \; \mathrm{cm}$ meteors.

Finally, we note that if gram-scale relativistic spacecraft such as the proposed Breakthrough Starshot\footnote{https://breakthroughinitiatives.org/initiative/3} project arrive to Earth from other civilizations \citep{Loeb2020} and come into contact with the Earth's atmosphere, they would appear as sub-relativistic $\sim 1 \mathrm{\; cm}$ meteors.

\section*{Acknowledgements}
This work was supported in part by a grant from the Breakthrough Prize Foundation.






\clearpage




\begin{thebibliography}{}
\makeatletter
\relax
\def\mn@urlcharsother{\let\do\@makeother \do\$\do\&\do\#\do\^\do\_\do\%\do\~}
\def\mn@doi{\begingroup\mn@urlcharsother \@ifnextchar [ {\mn@doi@}
  {\mn@doi@[]}}
\def\mn@doi@[#1]#2{\def\@tempa{#1}\ifx\@tempa\@empty \href
  {http://dx.doi.org/#2} {doi:#2}\else \href {http://dx.doi.org/#2} {#1}\fi
  \endgroup}
\def\mn@eprint#1#2{\mn@eprint@#1:#2::\@nil}
\def\mn@eprint@arXiv#1{\href {http://arxiv.org/abs/#1} {{\tt arXiv:#1}}}
\def\mn@eprint@dblp#1{\href {http://dblp.uni-trier.de/rec/bibtex/#1.xml}
  {dblp:#1}}
\def\mn@eprint@#1:#2:#3:#4\@nil{\def\@tempa {#1}\def\@tempb {#2}\def\@tempc
  {#3}\ifx \@tempc \@empty \let \@tempc \@tempb \let \@tempb \@tempa \fi \ifx
  \@tempb \@empty \def\@tempb {arXiv}\fi \@ifundefined
  {mn@eprint@\@tempb}{\@tempb:\@tempc}{\expandafter \expandafter \csname
  mn@eprint@\@tempb\endcsname \expandafter{\@tempc}}}
  
\bibitem[Berezinsky \& Prilutsky(1973)]{1973Ap&SS..21..475B} Berezinsky, V.~S., \& Prilutsky, O.~F.\ 1973, \apss, 21, 475

\bibitem[Berezinskii \& Prilutskii(1977)]{1977ICRC....2..358B} Berezinskii, V.~S., \& Prilutskii, O.~F.\ 1977, International Cosmic Ray Conference, 358

\bibitem[Bialy \& Loeb(2018)]{2018ApJ...868L...1B} Bialy, S., \& Loeb, A.\ 2018, \apjl, 868, L1

\bibitem[Bingham \& Tsytovich(1999)]{1999APh....12...35B} Bingham, R., \& Tsytovich, V.~N.\ 1999, Astroparticle Physics, 12, 35

\bibitem[Cherchneff \& Dwek(2009)]{2009ApJ...703..642C} Cherchneff, I., \& Dwek, E.\ 2009, \apj, 703, 642

\bibitem[Ceplecha et al.(1998)]{1998SSRv...84..327C} Ceplecha, Z., Borovi{\v{c}}ka, J., Elford, W.~G., et al.\ 1998, \ssr, 84, 327

\bibitem[Elenskii \& Suvorov(1977)]{1977Ap.....13..432E} Elenskii, Y.~S., \& Suvorov, A.~L.\ 1977, Astrophysics, 13, 432

\bibitem[Feige et al.(2012)]{2012PASA...29..109F} Feige, J., Wallner, A., Winkler, S.~R., et al.\ 2012, \pasa, 29, 109

\bibitem[Fimiani et al.(2016)]{2016PhRvL.116o1104F} Fimiani, L., Cook, D.~L., Faestermann, T., et al.\ 2016, \prl, 116, 151104

\bibitem[Herlofson(1956)]{1956Tell....8..268H} Herlofson, N.\ 1956, Tellus, 8, 268

\bibitem[Hayakawa(1972)]{1972Ap&SS..16..238H} Hayakawa, S.\ 1972, \apss, 16, 238

\bibitem[Hoang et al.(2015)]{2015ApJ...806..255H} Hoang, T., Lazarian, A., \& Schlickeiser, R.\ 2015, \apj, 806, 255

\bibitem[Hoang et al.(2017)]{2017ApJ...837....5H} Hoang, T., Lazarian, A., Burkhart, B., et al.\ 2017, \apj, 837, 5

\bibitem[Kachelrie{\ss} et al.(2015)]{2015PhRvL.115r1103K} Kachelrie{\ss}, M., Neronov, A., \& Semikoz, D.~V.\ 2015, \prl, 115, 181103

\bibitem[Kachelrie{\ss} et al.(2018)]{2018PhRvD..97f3011K} Kachelrie{\ss}, M., Neronov, A., \& Semikoz, D.~V.\ 2018, \prd, 97, 063011

\bibitem[Knie et al.(1999)]{1999PhRvL..83...18K} Knie, K., Korschinek, G., Faestermann, T., et al.\ 1999, \prl, 83, 18

\bibitem[Knie et al.(2004)]{2004PhRvL..93q1103K} Knie, K., Korschinek, G., Faestermann, T., et al.\ 2004, \prl, 93, 171103

\bibitem[Le Pichon et al.(2008)]{2008JGRD..11312115L} Le Pichon, A., Vergoz, J., Herry, P., et al.\ 2008, Journal of Geophysical Research (Atmospheres), 113, D12115

\bibitem[Loeb(2020)]{Loeb2020} Loeb, A.\ 2020, Surfing a Supernova, Scientific American, Observations

\bibitem[McBreen et al.(1993)]{1993Ap&SS.205..355M} McBreen, B., Plunkett, S., \& Lambert, C.~J.\ 1993, \apss, 205, 355

\bibitem[Richardson(2019)]{Richardson2019} Richardson, A S.\ 2019, NRL Plasma Formulary. Naval Research Laboratory, Washington, DC

\bibitem[Ruderman(1974)]{1974Sci...184.1079R} Ruderman, M.~A.\ 1974, Science, 184, 1079

\bibitem[Rybicki \& Lightman(1979)]{1979rpa..book.....R} Rybicki, G.~B., \& Lightman, A.~P.\ 1979, A Wiley-Interscience Publication

\bibitem[Sarazin et al.(2019)]{2019BAAS...51c..93S} Sarazin, F., Anchordoqui, L., Beatty, J., et al.\ 2019, \baas, 51, 93

\bibitem[Schanne et al.(2007)]{2007ESASP.622..117S} Schanne, S., Cass{\'e}, M., Sizun, P., et al.\ 2007, The Obscured Universe. Proceedings of the VI INTEGRAL Workshop, 117

\bibitem[Siraj \& Loeb(2019)]{2019arXiv190603270S} Siraj, A., \& Loeb, A.\ 2019, arXiv e-prints, arXiv:1906.03270

\bibitem[Spitzer(1949)]{1949PhRv...76..583S} Spitzer, L.\ 1949, Physical Review, 76, 583

\bibitem[Wallner et al.(2016)]{2016Natur.532...69W} Wallner, A., Feige, J., Kinoshita, N., et al.\ 2016, \nat, 532, 69

\bibitem[Wang \& Chevalier(2002)]{2002ApJ...574..155W} Wang, C.-Y., \& Chevalier, R.~A.\ 2002, \apj, 574, 155

\bibitem[Weiler(2003)]{Weiler2003} Weiler, K. W.\ 2003, \mn@doi [Supernovae and Gamma-Ray Bursters.] {10.1007/3-540-45863-8} Springer, Berlin, Heidelberg

\makeatother
\end{thebibliography}
\end{document}